\documentclass[aps,prx,reprint,superscriptaddress,showpacs]{revtex4-1}
\usepackage{graphicx}
\UseRawInputEncoding 
\usepackage{amssymb}
\usepackage{color}
\usepackage{amsmath}
\usepackage{ulem}
\usepackage{mathrsfs}

\newcommand{\up}{\uparrow}
\newcommand{\dn}{\downarrow}

\newcommand{\be}{\begin{eqnarray}}
\newcommand{\ee}{\end{eqnarray}}
\newcommand{\la}{\langle}
\newcommand{\ra}{\rangle}

\begin{document}

\title{Decoupling the Magnetic Field Operator as an Independent Operator Class in Stochastic Series Expansion Quantum Monte Carlo}
\author{Shijie Jin}
\affiliation{School of Physics, Beijing Institute of Technology, Beijing 100081, China}
\author{Lu Liu}
\email{liulu96@bit.edu.cn}
\affiliation{School of Physics, Beijing Institute of Technology, Beijing 100081, China}

\begin{abstract}   

The Stochastic Series Expansion (SSE) quantum Monte Carlo method with loop updates is among the most powerful approaches for quantum spin and boson systems. In the standard formulation, the magnetic field term is routinely absorbed into the Heisenberg interactions---a strategy that has proven highly efficient across a wide range of field strengths.
In this work, we propose a more general SSE framework in which the magnetic field operator is treated as an independent operator class, enabling it to participate in Monte Carlo updates on an equal footing with all other operators. Importantly, the field operator and other interaction operators can transform into one another during the update process. We demonstrate this algorithm using the two-dimensional antiferromagnetic Heisenberg model in a magnetic field as a concrete example. The simulation results agree with those of the conventional algorithm, confirming the correctness of the new formulation. A comparison of the integrated autocorrelation time shows that the new algorithm is competitive. This flexibility not only facilitates measurements of observables associated with operators but also offers potential computational advantages for a broader class of problems.
\end{abstract}

\date{\today}

\maketitle

\section{Introduction}
\label{sec:intro}

The Stochastic Series Expansion (SSE) quantum Monte Carlo (QMC) method with loop updates~\cite{sse1,sse2,sse-loop,sse-direct,sse-direct-2,alet2005} is a well-established and highly versatile technique for quantum spin~\cite{henelius2000,wessel2001,henelius2002,schmidt2008,alet2016,iaizzi2017,iaizzi2018,cuiyi} and boson~\cite{dorneich2002,schmid2002,wang2016,hesselmann2016} systems. Among its many applications, the Heisenberg model and its generalizations have received extensive attention. When an external magnetic field is present, Sylju\r{a}sen and Sandvik developed an SSE method with directed-loop updates that incorporate the field operators into the Heisenberg interactions~\cite{sse-direct,sse-direct-2}. Using the Heisenberg model as a benchmark, they showed that the directed-loop simulations remain highly efficient across the entire field range, from zero field up to saturation. Nevertheless, this merging strategy, while effective, is by no means the only possible route. For instance, Ref.~\cite{luliu2024} demonstrates that combining the magnetic field operator with other operators can lead to further efficiency gains. Recently, Ref.~\cite{dao2026} introduced a new algorithm specifically for the staggered magnetic field, in which the staggered-field operator is treated independently and updated through its own dynamics.

In this paper, taking the antiferromagnetic Heisenberg model with a magnetic field on a two-dimensional square lattice as an example, we argue that the magnetic-field operator possesses more flexibility than has been previously recognized. Not only can it be combined with or separated from other operators, but it can also undergo mutual transformations with other operators during the Monte Carlo evolution. To realize this idea, we develop a new SSE algorithm in which the magnetic field operator is released from its traditional subordinate role and allowed to interact with other operators on an equal footing. In contrast to previous schemes, our method is more general and enables the field operator and other interaction channels to be updated on the same footing and to transform into one another, thus offering a unified and more flexible framework.

This paper is organized as follows. In Sec.~\ref{sec:heisenbergmodel}, we introduce the antiferromagnetic Heisenberg model in a magnetic field. In Sec.~\ref{sec:newsse}, we present the new SSE algorithm in which the magnetic field is treated as an independent operator. In Sec.~\ref{sec:equation}, we derive the directed-loop equations for this algorithm. In Sec.~\ref{sec:simu}, we present simulation results that validate the correctness of the new SSE formulation and compare its efficiency with that of the conventional algorithm. Finally, we conclude in Sec.~\ref{sec:summary} with a summary and discussion.

\section{The Heisenberg model with magnetic field}
\label{sec:heisenbergmodel}
  The Hamiltonian of the antiferromagnetic Heisenberg model with magnetic field is expressed as
   \be
   H&=&-J\sum_{\la ij\ra}P_{ij}-h\sum_{i=1}^{N}S_i^z
   \ee
 where $P_{ij}=1/4-{\bf S_i}\cdot{\bf S_j}$ is the singlet projector operator, and $\la ij\ra$ denotes two nearest neighbor sites. The summations run over all nearest neighbors for the $J$ terms and over all $N$ lattice sites for the $h$ term. In this paper, we set $J=1$ as the unit of energy. In the conventional SSE formulation, the Hamiltonian is divided into two different types of bond operators:
 \be
 H_{1,b}&=&1/4-S_{i(b)}^zS_{j(b)}^z+\frac{h}{z}(S_{i(b)}^z+S_{j(b)}^z+1+\epsilon),\nonumber\\
 H_{2,b}&=&\frac{1}{2}(S_{i(b)}^+S_{j(b)}^-+S_{i(b)}^-S_{j(b)}^+)
 \ee
where $H_{1,b}$ represents the diagonal operator and $H_{2,b}$ denotes the off-diagonal operator. $\epsilon$ is a nonnegative constant that does not affect the physical properties of the system, and $z$ denotes the coordination number.
With this decomposition, the SSE method with directed-loop updates can be directly applied~\cite{sse-direct,sse-direct-2}.

This conventional treatment, which merges the magnetic field operator with the diagonal part of the Heisenberg operator, inevitably mixes the two components and makes them difficult to distinguish. In contrast to this fixed combination, the magnetic field operator is by nature an on-site diagonal operator, and can therefore be flexibly associated with diagonal operators acting on multiple lattice sites.
Such flexibility allows for different groupings, giving rise to various combined diagonal operators and enabling the field operator to be merged with different types of operators-for instance, with the Heisenberg diagonal operator or with the multi-spin interaction operators explored in Ref.~\cite{luliu2024}. In this paper, we pursue an alternative strategy: instead of merging the field operator with others, we treat it as an independent operator class that can be updated on its own and, more importantly, can mutually transform with other operators during the Monte Carlo update.

Since the Heisenberg bond operators act on two sites, we recast the magnetic field operator into a two-site bond operator form as well. We divide the Hamiltonian into three different bond operators:
\be
 H_{1,b}&=&1/4-S_{i(b)}^zS_{j(b)}^z,\nonumber\\
 H_{2,b}&=&\frac{1}{2}(S_{i(b)}^+S_{j(b)}^-+S_{i(b)}^-S_{j(b)}^+),\nonumber\\
 H_{3,b}&=&\frac{h}{z}(S_{i(b)}^z+S_{j(b)}^z+1+\epsilon).
 \label{eq:newbond}
\ee
where $H_{1,b}$ denotes the diagonal Heisenberg operator, $H_{2,b}$ denotes the off-diagonal Heisenberg operator, and $H_{3,b}$ is the diagonal magnetic field operator. 

\begin{figure}[t]
  \includegraphics[width=75mm,clip]{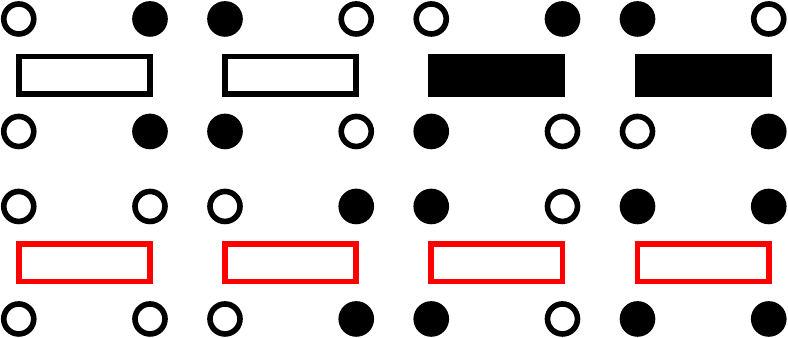}
  \caption{Vertex representations of the three types of bond operators. The open circles represent spin-up ($\up$) states, and the solid circles represent  spin-down ($\dn$) states. The horizontal bars represent the operators. The first row shows the four vertices allowed for the Heisenberg operators, where the first two black empty bond configurations correspond to the diagonal Heisenberg operators and the last two black filled bond configurations correspond to the off-diagonal Heisenberg operators. The second row displays the four bond configurations for the diagonal magnetic-field operators, represented by red empty bonds.
  We label these eight vertices as $\Gamma_i$, ($i=1,2,\cdots,8$) and denote the four spins as the four legs of the vertex.}
  \label{fig:vertex}
  \end{figure}

With this representation, we construct a Monte Carlo update scheme in which the three types of operators in Eq.~(\ref{eq:newbond}) can be transformed into one another. The vertex representations of these three operator types are illustrated in Fig.~\ref{fig:vertex}. The first row shows the four vertices allowed for the Heisenberg operators, where the first two black empty bond configurations correspond to the diagonal Heisenberg operators $H_{1,b}$ and the last two black filled bond configurations correspond to the off-diagonal Heisenberg operators $H_{2,b}$. The second row displays the four bond configurations for the diagonal magnetic-field operators $H_{3,b}$, which are represented by red empty bonds. 

The matrix elements (weight) of these eight vertices are
\be
\Gamma_1:\la\up\dn|H_{1,b}|\up\dn\ra&=&1/2,\nonumber\\
\Gamma_2:\la\dn\up|H_{1,b}|\dn\up\ra&=&1/2,\nonumber\\
\Gamma_3:\la\up\dn|H_{2,b}|\dn\up\ra&=&1/2,\nonumber\\
\Gamma_4:\la\dn\up|H_{2,b}|\up\dn\ra&=&1/2,\nonumber\\
\Gamma_5:\la\up\up|H_{3,b}|\up\up\ra&=&\frac{h}{z}(2+\epsilon),\nonumber\\ \Gamma_6:\la\up\dn|H_{3,b}|\up\dn\ra&=&\frac{h}{z}(1+\epsilon),\nonumber\\
\Gamma_7:\la\dn\up|H_{3,b}|\dn\up\ra&=&\frac{h}{z}(1+\epsilon),\nonumber\\
\Gamma_8:\la\dn\dn|H_{3,b}|\dn\dn\ra&=&\frac{h}{z}\epsilon.
\label{eq:matrix}
\ee

For simplicity, we set $\epsilon=0$ in this work. We note that the algorithm does not rely on this specific choice, and other nonnegative values of $\epsilon$ are equally applicable. With this choice, the Hamiltonian becomes:
\be
H=-\sum_{b=1}^{N_b}(H_{1,b}-H_{2,b}+H_{3,b})+c.
\label{eq:hamiltonian}
\ee
 Each of the three operator types has $N_b=2L^2$ bonds on the two-dimensional square lattice, and the constant $c=Nh/2$ in this case, which is neglected in the simulations.

\section{New SSE algorithm with magnetic field}
\label{sec:newsse}
Based on the framework established in the previous section, we describe the new SSE algorithm in detail here. We work in the standard basis for Heisenberg models with $N$ spins:
\be
|\alpha\ra=|S_1^z,S_2^z,\cdots,S_N^z\ra.
\ee

The starting point of the SSE method is the Taylor expansion of the partition function:
\be
Z={\rm Tr} \{{\rm e}^{-\beta H}\}=\sum_\alpha\sum_{n=0}^\infty \frac{(-\beta)^n}{n!}\la\alpha|H^n|\alpha\ra
\ee
where the trace is expressed as a summation over the basis ${|\alpha\ra}$ and $\beta=1/T$ is the inverse temperature. According to Eq.~(\ref{eq:hamiltonian}), the operator string $H^n$ expands into a sum over products of bond operators. We truncate the expansion order $n$ at a maximum value $M$ and then fix the operator products length to $M$ by introducing unit operators. For a given expansion power $n$ $(n\leq M)$, we insert $M-n$ unit operators $H_{0,0}=I$ into the operator products in all possible ways. Finally, the partition function takes the form
\be
Z=\sum_\alpha\sum_{S_M}(-1)^{n_2}\frac{\beta^n(M-n)!}{M!}\la \alpha|\prod_{p=1}^M H_{a(p),b(p)}|\alpha\ra
\label{part}
\ee
where $n$ denotes the number of non-unit bond operators, $n_2$ is the number of $H_{2,b}$ operators, $a(p)=0,1,2,3$ corresponds to the operator type  
and $b(p)=0,1,2,\cdots,N_b$ is the bond index ($0$ for unit operators, $1,2,\cdots,N_b$ for non-unit bond operators). The sign factor $(-1)^{n_2}$ is positive for bipartite lattice~\cite{sse-loop}.
$S_M$ is the configurations of operator products.
Such a product can be represented by an operator-index sequence
\be
S_M=[a(1),b(1)],[a(2),b(2)],\cdots,[a(M),b(M)].
\ee
For simplicity, we sometimes write $[a,b]_p$ to denote $[a(p),b(p)]$, where $p$ can be interpreted as the index of imaginary time.

One can easily show that the average expansion order is
\be
\la n\ra=\beta|E|
\ee
where $E$ is the system energy, $E=\la H\ra$~\cite{sse-loop,sse-direct}. The fluctuation of the expansion order is approximately $\la n\ra^{1/2}$. The cutoff $M$ can therefore be chosen large enough that $n$ never reaches this limit, so that the truncation error is negligible.

The Monte Carlo simulation can be initialized with a random state $|\alpha\ra$ and unit-operator string $S_M=[0,0]_1,[0,0]_2,\cdots,[0,0]_M$. The SSE sampling of configurations $(\alpha, S_M)$ involves two distinct types of updates, which together ensure the ergodicity of the sampling.

\begin{figure}[t]
  \includegraphics[width=80mm,clip]{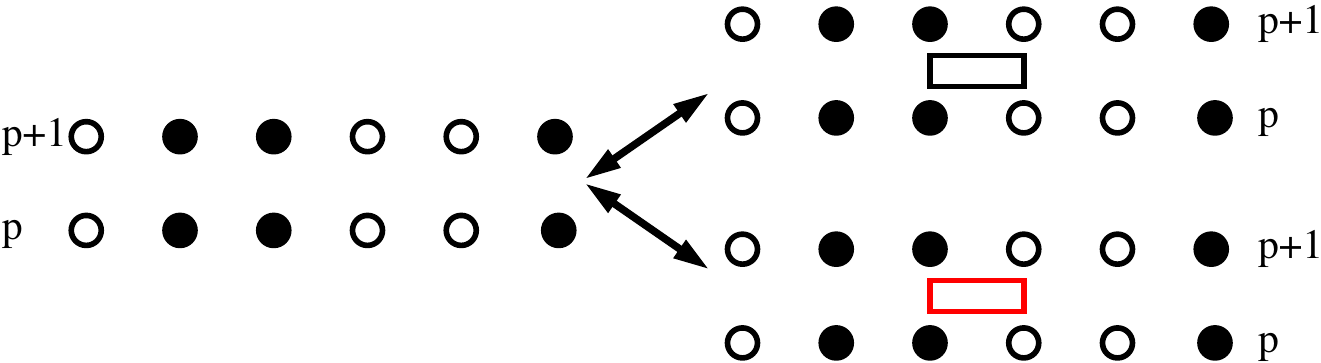}
  \caption{Diagonal update. A diagonal operator is either inserted, $[0,0]_p\rightarrow[i,b]_p$, at a randomly selected bond or removed, $[i,b]_p\rightarrow[0,0]_p$.}
  \label{fig:diagonal}
  \end{figure}

The first type of update, the diagonal update, involves the update between the unit operator $[0,0]_p$ and the diagonal operator $[i,b]_p~(i=1$ or $3$). Such an update changes the expansion order $n$ by $\pm 1$. This update scheme is illustrated in Fig.~\ref{fig:diagonal}. If the operator at imaginary-time $p$ is the unit one, a diagonal bond is randomly chosen from $H_{1,b}$ and $H_{3,b}$. This bond is attempted to insert into the configuration with a probability. If the operator is a diagonal one, it is removed with a probability. The corresponding Metropolis acceptance probabilities for inserting a diagonal operator and for removing one are
 \be
 P([0,0]_p\rightarrow[i,b]_p)={\rm min}\left\{1,\frac{2N_b\beta\la\alpha(p)|H_{i,b}|\alpha(p)\ra}{M-n}\right\},\nonumber\\
 P([i,b]_p\rightarrow[0,0]_p)={\rm min}\left\{1,\frac{M-n+1}{2N_b\beta\la\alpha(p)|H_{i,b}|\alpha(p)\ra}\right\},\nonumber
 \ee
 where the matrix elements are given in Eq.~(\ref{eq:matrix}). These diagonal updates are attempted sequentially for all $p=1,2,\cdots,M$. 

\begin{figure}[ht]
  \includegraphics[width=80mm,clip]{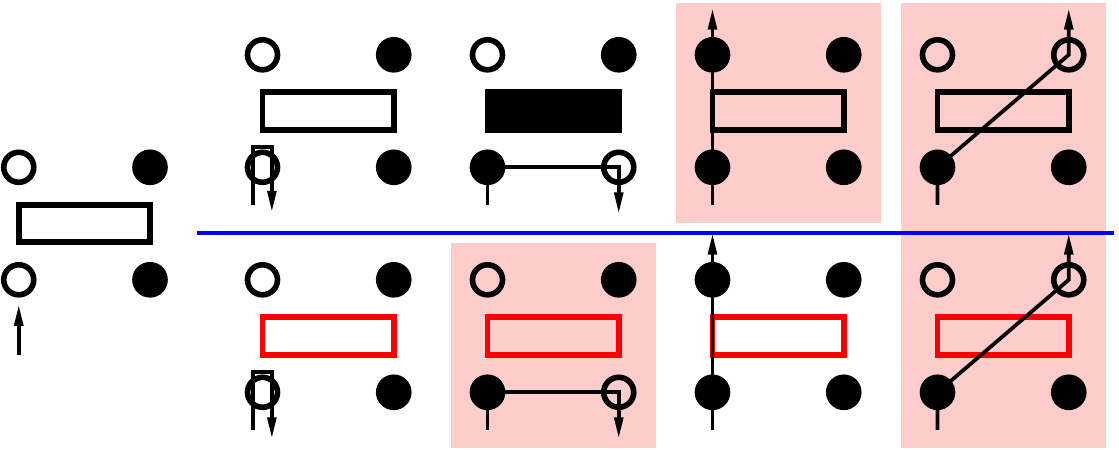}
  \caption{All eight possible update choices for vertex $\Gamma_1$ with the entrance located at the lower left. The arrow indicates the exit leg. The resulting updated vertices are shown on the right, where the Heisenberg operator vertices are displayed above the blue line and the magnetic-field operator vertices below the blue line. The four vertices in the pink region are forbidden.}
  \label{fig:update-example}
  \end{figure}

The second type of update, the off-diagonal update, involves the update between the diagonal operators $[i,b]_p~(i=1$ or $3)$ and the off-diagonal operators $[2,b]_p$. This is achieved by the directed-loop update~\cite{sse-loop,sse-direct}. To construct a loop, one leg of a vertex is randomly selected as an initial entrance leg. An exit leg is then chosen among the four legs of the same vertex, after which both the entrance and exit spins are flipped. In addition to the spin configuration of the vertex, the bond types at this vertex should also be changed. In the conventional scheme, the bond type can only change from $H_{1,b}$ and $H_{2,b}$. In our algorithm, however, updates among all three bond types are allowed. Examples of vertex update processes are shown in Fig.~\ref{fig:update-example}. There are eight possible update choices, only four are permitted. The probability of exiting at a given leg with a certain bond type is determined by the directed-loop equations, which will be presented in the next section. The leg to which the exit leg is connected then becomes the entrance leg of the next vertex. A new exit leg and bond type are again chosen for this new vertex. This procedure continues until the loop returns to the original starting point,  at which the loop is closed. Such a closed loop update generates a new configuration that contributes to the partition function. If a spin is not acted upon by any operators, it is flipped with a probability $1/2$.

It is useful to introduce a Monte Carlo step (MCS) for the SSE simulation. A single MCS consists of a complete diagonal-update sweep across all imaginary-time indices, subsequently the linked-list construction~\cite{sse-loop}, and finally a fixed number of loop updates. In this way, both update types are included in every MCS.

During the simulation, the initial configuration must first be evolved to the equilibrium configurations before reliable measurements can be performed. Consequently, in Monte Carlo simulations, one performs a number of ``equilibration" MCSs followed by a number of ``measurement" MCSs. The required numbers of these two types of MCSs are determined by the equilibrium correlation time and the autocorrelation time; a detailed discussion of this issue is beyond the scope of the present work. Physical observables are measured during the ``measurement" MCSs. A general observable $A$ (mostly diagonal) can be measured according to
\be
\la A\ra&=&\frac{1}{Z}\sum_{\alpha,S_M}\frac{\beta^n(M-n)!}{M!}\la \alpha|A\prod_{p=1}^M H_{a(p),b(p)}|\alpha\ra\nonumber\\
&=&\sum_{\alpha,S_M}A(\alpha,S_M)W(\alpha,S_M)~/\sum_{\alpha,S_M}W(\alpha,S_M)\nonumber
\ee
where 
\be
A(\alpha,S_M)&=&\frac{\la \alpha|A\prod_{p=1}^M H_{a(p),b(p)}|\alpha\ra}{\la \alpha|\prod_{p=1}^M H_{a(p),b(p)}|\alpha\ra}\nonumber\\
W(\alpha,S_M)&=&\frac{\beta^n(M-n)!}{M!}\la \alpha|\prod_{p=1}^M H_{a(p),b(p)}|\alpha\ra.\nonumber
\ee

We do not discuss the measurement details in depth here, as several observables have already been derived in Refs.~\cite{sse-loop,obs}, and off-diagonal correlation functions have been studied in Ref.~\cite{obs-off}.

\section{Directed-loop Equation}
\label{sec:equation}
As discussed previously, the probability of exiting a vertex at a given leg with a given bond type is determined by the directed-loop equations, which we introduce and solve in this section.

The directed-loop equations have been discussed extensively in Refs.~\cite{sse-direct,sse-direct-2}. In this work, we only present a brief introduction to them.
To maintain the detailed balance in the directed-loop update, any vertex in the loops should satisfy:
\be
W(s,e;s^\prime,x)&=&W(s^\prime,x;s,e),\nonumber\\
\sum_xW(s,e;s^\prime,x)&=&W_s,\label{eq:set}
\ee
where $s$ denotes the vertex configuration with weight $W_s$. $W(s,e;s^\prime,x)\equiv W_sP(s,e\rightarrow s^\prime,x)$, with $e$ and $x$ being the entrance and exit legs, respectively. Given a vertex with configuration $s$ and entrance leg $e$, the quantity $P(s,e\rightarrow s^\prime,x)$ is the probability that the loop exits at leg $x$ with a new vertex configuration $s^\prime$. 

\begin{figure}[h]
  \includegraphics[width=80mm,clip]{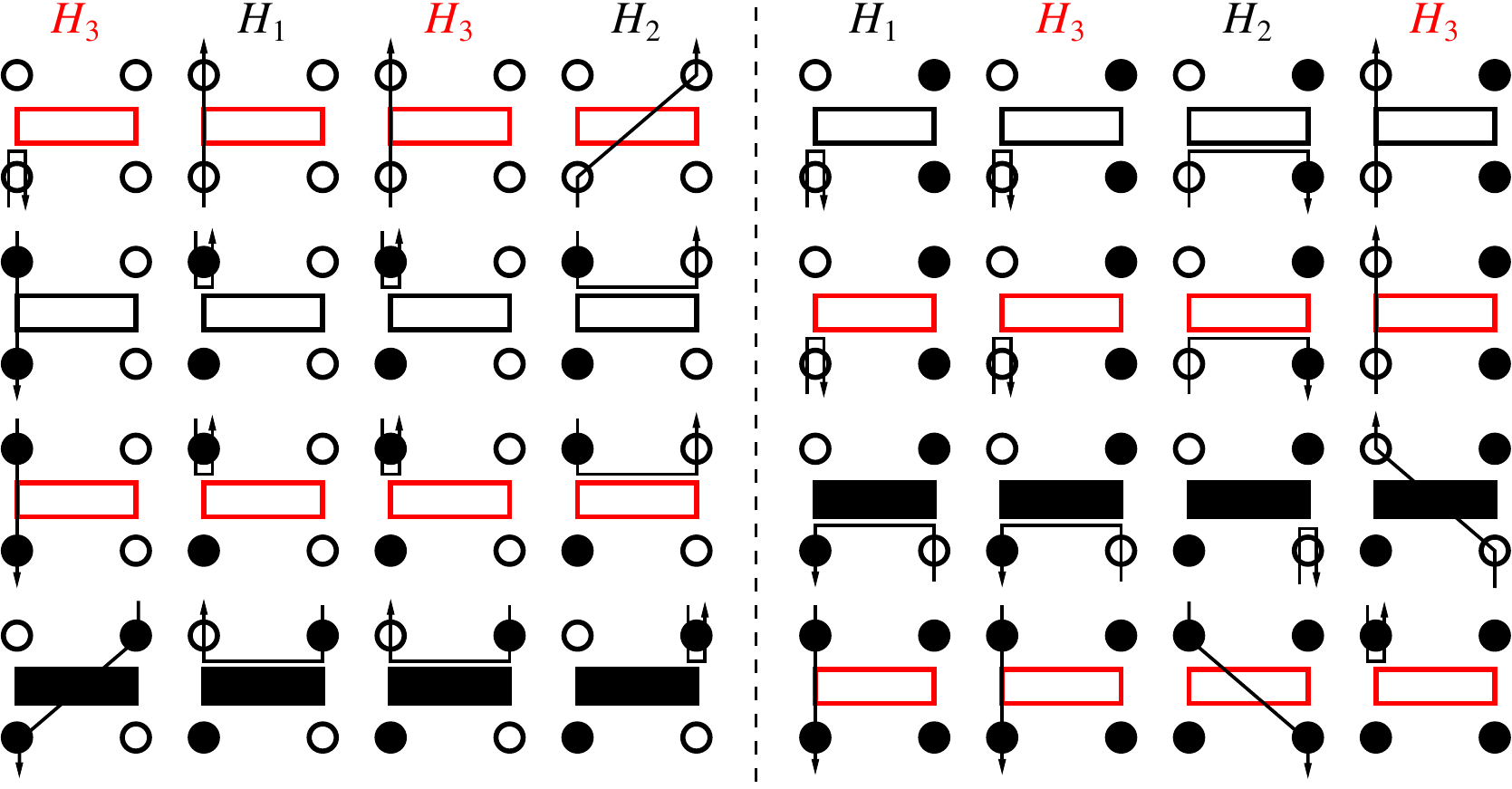}
  \caption{Two independent segments of directed-loop assignments. All other assignments can be derived from these two segments by certain symmetry transformations. The vertices represent the vertices before the update. The lines with arrows represent the directed loops, and the arrows indicate the loop directions. For each column, the bond type of the updated vertex is shown at the top, where $H_i$ ($i=1,2,3$) denote the three different bond operators defined in Eq.~(\ref{eq:newbond}).}
  \label{fig:twoset}
  \end{figure}

For a given vertex and entrance leg, there are four possible vertex update choices. Across all vertex types and all possible entrance legs, the resulting update processes can be classified into two independent segments of assignments, as shown in Fig.~\ref{fig:twoset}. 
These two segments correspond to the two topologically distinct vertex-update sectors. All others can be obtained from these two segments by symmetry transformations. 

From Eq.~(\ref{eq:set}), the directed-loop equations for each set can be derived, and their solutions are readily obtained. It should be noted that each set of equations admits an infinite number of solutions. The particular solution adopted in this work is chosen to minimize the probability of the ``no-update" event-i.e., the case where the exit leg is the same as the entrance leg and the bond type remains unchanged. Such treatment is also used in Refs.~\cite{sse-direct,sse-direct-2}. This choice is based on the intuitive expectation that suppressing such event may improve the simulation efficiency. However, it cannot be excluded that a more efficient directed-loop solution exists in which this probability is not minimized.

The two independent sets of equations corresponding to the two segments in Fig.~\ref{fig:twoset} are given by
\be
W_1&=&b_1+a+b+c,\nonumber\\
W_2&=&a+b_2+d+e,\nonumber\\
W_3&=&b+d+b_3+f,\nonumber\\
W_4&=&c+e+f+b_4,
\label{eq:solu1}
\ee
for the left set, and
\be
W_5&=&b_1^\prime+a^\prime+b^\prime+c^\prime,\nonumber\\
W_6&=&a^\prime+b_2^\prime+d^\prime+e^\prime,\nonumber\\
W_7&=&b^\prime+d^\prime+b_3^\prime+f^\prime,\nonumber\\
W_8&=&c^\prime+e^\prime+f^\prime+b_4^\prime,
\label{eq:solu2}
\ee
for the right set. On the left-hand sides, the symbols $W_i$ denote the vertex weights, which are given in Eqs.~(\ref{eq:matrix}):

\be
W_1&=&\Gamma_5=\frac{h}{z}(2+\epsilon),\quad W_2=\Gamma_2=1/2,\nonumber\\
W_3&=&\Gamma_7=\frac{h}{z}(1+\epsilon),\quad W_4=\Gamma_3=1/2,\nonumber\\
W_5&=&\Gamma_1=1/2,\qquad W_6=\Gamma_6=\frac{h}{z}(1+\epsilon),\nonumber\\
W_7&=&\Gamma_3=1/2,\qquad W_8=\Gamma_8=\frac{h}{z}\epsilon,
\ee
with $\epsilon=0$ in this work. On the right-hand sides, the terms are the corresponding $W(s,e;s^\prime,x)\equiv W_sP(s,e\rightarrow s^\prime,x)$, and the probability of choosing the new vertex configuration $s^\prime$ is $P(s,e\rightarrow s^\prime,x)=W(s,e;s^\prime,x)/W(s)$. The solution of the $4\times4$ equation sets is discussed in detail in appendix. 
Based on these solutions, we implement the new directed-loop update into the Monte Carlo simulation. The corresponding results are presented in the next section.

\section{Simulation Results}
\label{sec:simu}
In this section, we present simulation results for our method. We first demonstrate the correctness of the algorithm, and then compare its efficiency with that of conventional methods.

To verify the correctness of our new algorithm, we compare the simulation results with those of the conventional method~\cite{sse-direct,sse-direct-2}. As our new algorithm focuses on the properties of models with magnetic field, we mainly concentrate on the magnetization, which is defined as
\be
M_z=\sum_{i=1}^NS_i^z,
\ee
and the energy density $e=E/N=-\la n\ra/\beta N$.

\begin{figure}[ht]
  \includegraphics[width=80mm,clip]{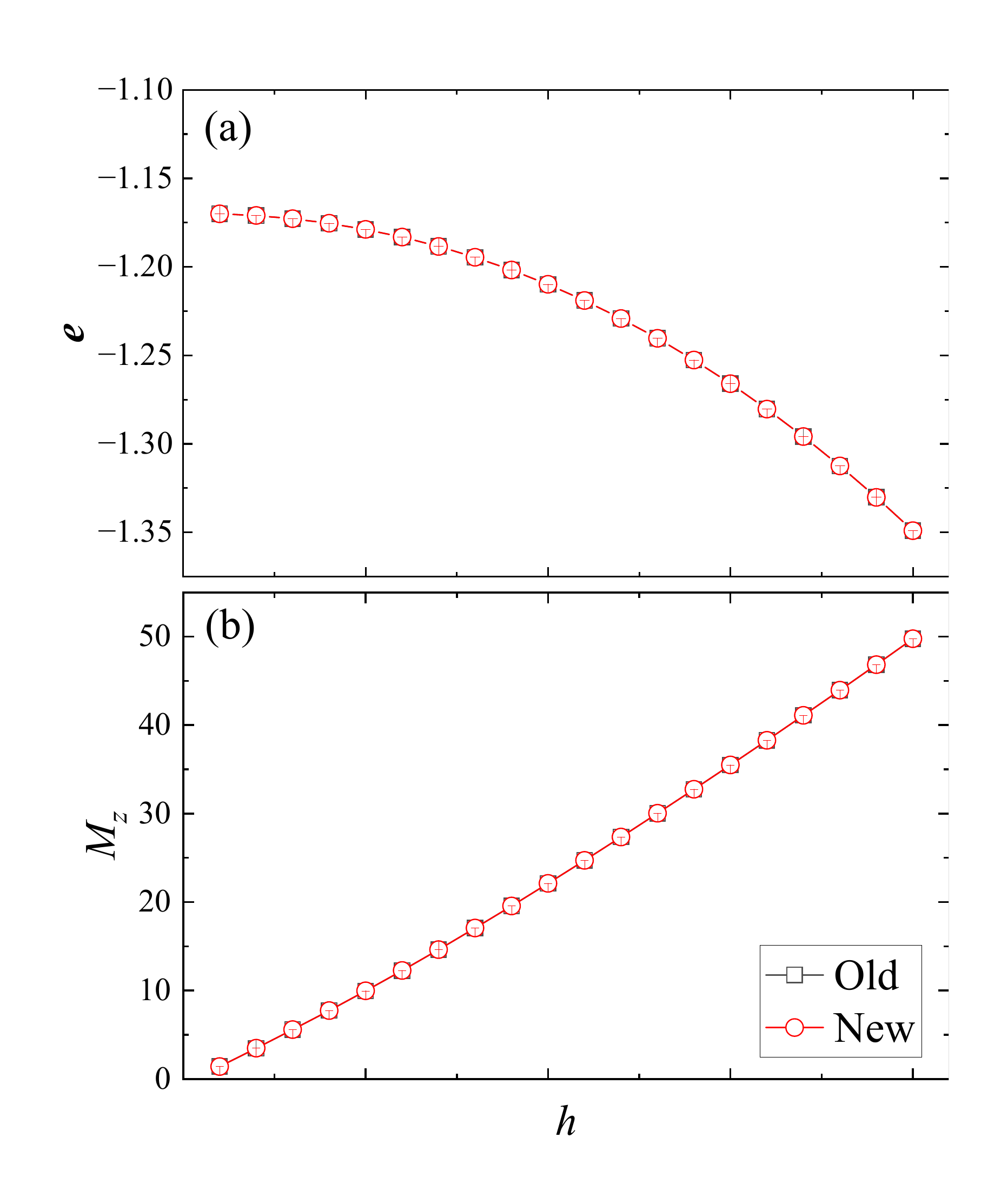}
  \caption{Energy density and total magnetization versus magnetic field for the Heisenberg model on a $16\times16$ square lattice at $\beta=16$, obtained from the conventional algorithm (Old, blue squares) and our new algorithm (New, red circles). }
  \label{fig:enmz}
  \end{figure}

In Fig.~\ref{fig:enmz}, we compare the energy density and total magnetization obtained from the conventional algorithm (Old) and from our new algorithm (New) for the Heisenberg model on a $16\times16$ square lattice at $\beta=16$. The results show good agreement.

The autocorrelation functions provide a direct quantitative measure of the efficiency of a 
Monte Carlo method in generating independent configurations. For a given observable $O$, the normalized autocorrelation function is defined as
\be
A_O(t)=\frac{\la O(i+t)O(i)\ra-\la O(i)\ra^2}{\la O(i)^2\ra-\la O(i)\ra^2},
\ee
where $i$ and $t$ denote Monte Carlo times, measured in units of $1$ MCS, and the angle brackets denote the average over time $i$. At large time separations, $A_O(t)$ decays exponentially,
\be
A_O(t)\xrightarrow{t\rightarrow\infty}a{\rm e}^{-t/\tau_{\mathrm{exp}}},
\label{eq:exp}
\ee
with $\tau_{\mathrm{exp}}$ the exponential autocorrelation time and $a$ a constant. This time is set by the slowest mode to which the observable $O$ couples; at smaller times, other modes contribute, and $O(t)$ no longer behaves purely exponentially.

The integrated autocorrelation time (IAT), defined as
\be
\tau_{\mathrm{int}}(O)=1/2+\sum_{t=1}^{\infty}A_O(t),
\label{eq:int}
\ee
provides a quantitative measure of the efficiency of the Monte Carlo algorithm~\cite{int-cor}. In this work, we focus on $\tau_{\mathrm{int}}$ of the total magnetization.

\begin{figure}[ht]
  \includegraphics[width=80mm,clip]{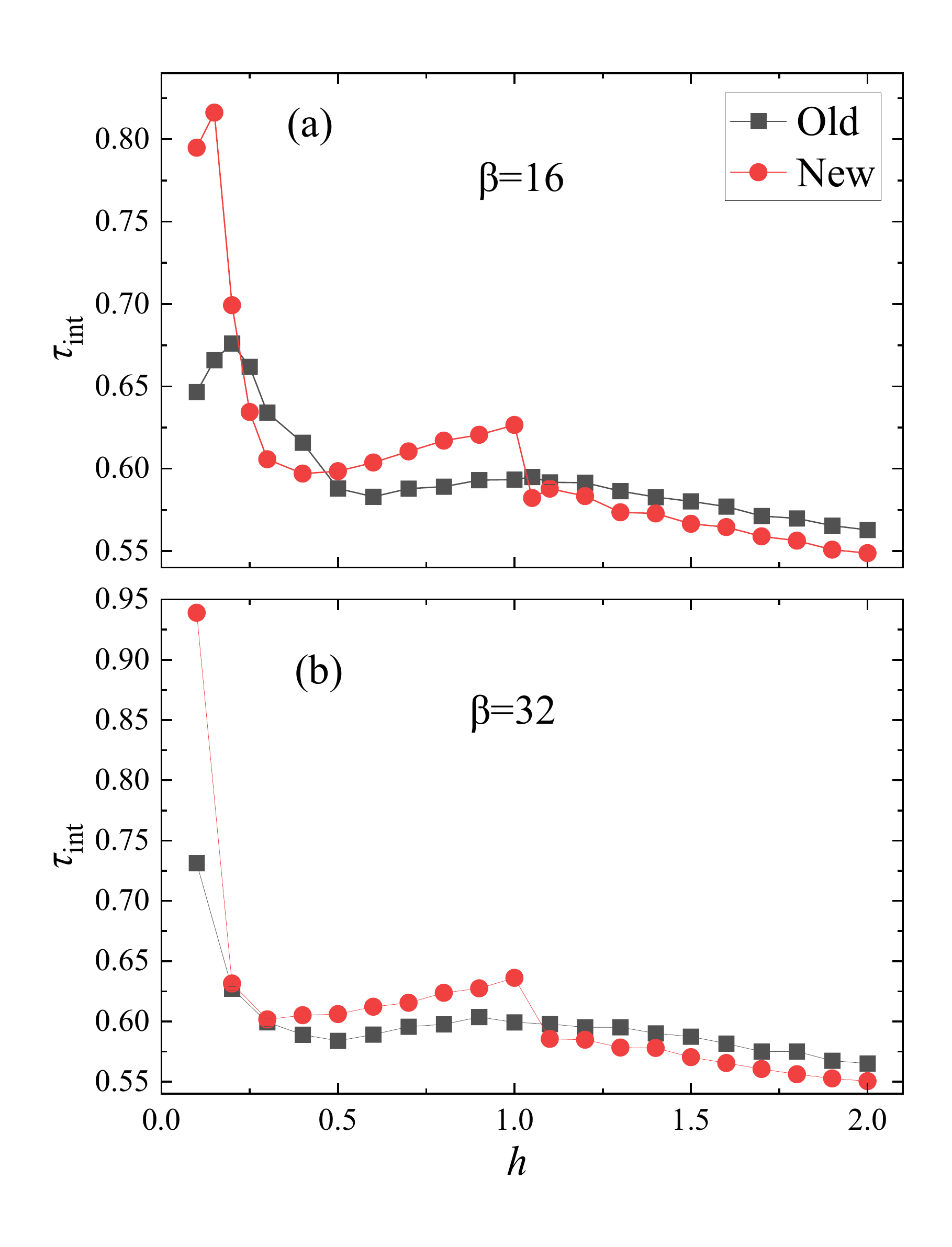}
  \caption{Integrated autocorrelation time of the total magnetization versus magnetic field for the Heisenberg model on a $16\times16$ square lattice at (a) $\beta=16$ and (b) $\beta=32$, obtained from the conventional algorithm (Old, blue squares) and our new algorithm (New, red circles). }
  \label{fig:tau1632}
  \end{figure}

Figure~\ref{fig:tau1632} shows the field dependence of the IAT for the magnetization with $L=16$ at $\beta=16$ and $32$. It can be seen that the IAT is small for both algorithms, indicating that both are efficient. For weak magnetic fields, the IAT of the New algorithm is slightly larger. However, as the magnetic field increases, the IAT of the New algorithm becomes smaller than that of the conventional one. We emphasize that, in our new algorithm, we simply take $\epsilon=0$ in Eqs.~(\ref{eq:matrix}). Any nonnegative value is allowed, and the IAT may be even smaller than in the $\epsilon=0$ case. We also observe a discontinuity in the IAT for our new algorithm. It arises because the directed-loop solution is piecewise in $h$: the solution in one field regime is not smoothly connected to that in the neighboring regime, so the parameters in Eqs.~(\ref{eq:solu1}) and (\ref{eq:solu2}) are not guaranteed to vary continuously.

\section{Summary and Discussion}
\label{sec:summary}
In this paper, we have proposed a generalized SSE quantum Monte Carlo framework in which the magnetic field operator is decoupled from the Heisenberg interactions and treated as an independent operator class. 
Taking the two-dimensional antiferromagnetic Heisenberg model in a magnetic field as a concrete example, we decomposed the Hamiltonian into three types of bond operators: the diagonal Heisenberg operator $H_{1,b}$, the off-diagonal Heisenberg operator $H_{2,b}$, and the diagonal magnetic-field operator $H_{3,b}$. Within this representation, the magnetic-field operator participates in the directed-loop updates on an equal footing with the Heisenberg operators and can mutually transform with them, in contrast to the conventional scheme in which the field is rigidly absorbed into the diagonal Heisenberg operator.

We derived the directed-loop equations for the new algorithm and solved them in the Appendix by minimizing the no-update probability. The correctness of the algorithm was verified by comparing the energy density and the total magnetization with those obtained from the conventional directed-loop method; the two approaches are in excellent agreement. The efficiency was benchmarked through the IAT of the magnetization. Both algorithms are efficient over the entire field range; while the new algorithm shows a slightly larger autocorrelation time in weak fields, it becomes more efficient than the conventional one as the field increases.

It should be emphasized that the present results correspond to a specific and simple choice of the free parameters, namely $\epsilon=0$ and a directed-loop solution that minimizes the no-update probability. The algorithm itself does not rely on these choices, and any nonnegative $\epsilon$ as well as other directed-loop solutions are equally admissible. These degrees of freedom provide additional room for optimization, and it is likely that a more efficient solution can be found that further reduces the autocorrelation time. The discontinuity observed in the IAT is a consequence of the particular piecewise solution adopted here and does not reflect a fundamental limitation of the algorithm.

The flexibility of the magnetic field operator uncovered in this work is expected to be of broader use. As an on-site diagonal operator, it can mutually update and transform with other operators acting on multiple sites, such as those for multi-spin interactions or staggered fields. Extending the present algorithm to such systems and exploring the associated computational advantages would be interesting directions for future work.

\begin{acknowledgments}
This work is supported by the National Natural Science Foundation of China under Grants No.12674288, 12304171 and Beijing Institute of Technology Research Fund Program for Young Scholars. 
\end{acknowledgments}
\appendix
\section*{Appendix}
In this appendix, we present the solution to the directed-loop equations used in this work. As stated in the main text, we set $\epsilon=0$ and minimize the probability of the no-update event for simplicity. Other choices of positive $\epsilon$ and alternative solutions are possible and may even be more efficient than the one adopted here; however, the underlying idea remains the same.

A general set of $4\times4$ equations takes the form
\be
W_1&=&a_{11}+a_{12}+a_{13}+a_{14}\nonumber\\
W_2&=&a_{21}+a_{22}+a_{23}+a_{24}\nonumber\\
W_3&=&a_{31}+a_{32}+a_{33}+a_{34}\nonumber\\
W_4&=&a_{41}+a_{42}+a_{43}+a_{44}
\ee
with $W_4\leq W_3\leq W_2\leq W_1$. In this set, all $a_{ij}$ should be non-negative and $a_{ij}=a_{ji}$. The probabilities of no-update events are determined by the diagonal elements $a_{ii}~(i=1,2,3,4)$, which are minimized in our solutions.

When $W_1<W_2+W_3+W_4$, it is possible to set all $a_{ii}=0$ and the solution used in this work is
\be
a_{11}=a_{22}=a_{33}=a_{44}=0,\nonumber\\
a_{12}=a_{21}=(W_1+W_2-W_3-W_4)/2,\nonumber\\
a_{13}=a_{31}=(W_1-W_2+W_3-W_4)/2,\nonumber\\
a_{14}=a_{41}=W_4,\nonumber\\
a_{23}=a_{32}=(-W_1+W_2+W_3+W_4)/2,\nonumber\\
a_{24}=a_{42}=a_{34}=a_{43}=0.
\label{eq:solution1}
\ee

However, when $W_1\ge W_2+W_3+W_4$, it is impossible to satisfy $a_{ii}=0$ for all $i$, and the solution used in this work is
\be
a_{11}=W_1-W_2-W_3-W_4,\nonumber\\
a_{22}=a_{33}=a_{44}=0,\nonumber\\
a_{12}=a_{21}=W_2,\nonumber\\
a_{13}=a_{31}=W_3,\nonumber\\
a_{14}=a_{41}=W_4,\nonumber\\
a_{23}=a_{32}=0,\nonumber\\
a_{24}=a_{42}=0,\nonumber\\
a_{34}=a_{43}=0.
\label{eq:solution2}
\ee

Based on these solutions, we now present the explicit results for Eqs.~(\ref{eq:solu1}) and (\ref{eq:solu2}). For simplicity, we set $h_b=h/z$.

For the solutions of Eqs.~(\ref{eq:solu1}), the results are given as follows.
In the regime $h_b\le1/4$
\be
b_1=b_2=b_3=b_4=0;\nonumber\\
a=h_b/2;~b=0;~c=3h_b/2;\nonumber\\
d=h_b;~e=1/2-3h_b/2;~f=0.
\ee

For $1/4 < h_b\le1/2$
\be
b_1=b_2=b_3=b_4=0;\nonumber\\
a=h_b/2;~b=h_b;~c=h_b/2;\nonumber\\
d=0;~e=1/2-h_b/2;~f=0.
\ee

For $1/2<h_b\le1$
 \be
b_1=b_2=b_3=b_4=0;\nonumber\\
a=h_b/2;~b=3h_b/2-1/2;~c=1/2;\nonumber\\
d=1/2-h_b/2;~e=0;~f=0.
\ee

For $h_b>1$
\be
b_1=h_b-1;~b_2=b_3=b_4=0;\nonumber\\
a=1/2;~b=h_b;~c=1/2;\nonumber\\
d=0;~e=0;~f=0.
\ee

For the solutions of Eqs.~(\ref{eq:solu2}), the results are:

For $h_b\le1$
\be
b_1^\prime=b_2^\prime=b_3^\prime=b_4^\prime=0;\nonumber\\
a^\prime=h_b/2;~b^\prime=1/2-h_b/2;~c^\prime=0;\nonumber\\
d^\prime=h_b/2;~e^\prime=0;~f^\prime=0.
\ee

For $h_b>1$
\be
b_2^\prime=h_b-1;~b_1^\prime=b_3^\prime=b_4^\prime=0,\nonumber\\
a^\prime=1/2;~b^\prime=0;~c^\prime=0,\nonumber\\
d^\prime=1/2;~e^\prime=0;~f^\prime=0.
\ee

We note that our solutions do not guarantee continuous variation of the parameters with $h_b$, so that the quantity $\tau_{\mathrm{int}}$ exhibits a discontinuity.

\end{document}